\documentclass[11pt]{article} 

\usepackage[margin=1in]{geometry}
\usepackage{natbib}
\usepackage{graphicx}
\usepackage{booktabs}
\usepackage{hyperref}
\usepackage{amsmath}

\usepackage{tikz}
\usetikzlibrary{arrows.meta,positioning,fit,calc}
\definecolor{accCol}{HTML}{1F6FB2}
\definecolor{tokCol}{HTML}{C62828}
\definecolor{neuCol}{HTML}{6E6E6E}

\title{LLM Evolution in a Simple Simulated Market}

\author{
  Charlie Pilgrim$^{1}$ \and 
  Christopher Krupenye$^{2}$ \and 
  Dora Biro$^{3}$ \and 
  Richard P. Mann$^{1}$ \\[6pt]
  $^{1}$University of Leeds \quad
  $^{2}$Johns Hopkins University \quad
  $^{3}$University of Rochester
}

\date{}

\begin{document}

\maketitle 

\begin{abstract}
Arguments that AI systems will evolve in harmful ways are largely made in the abstract. We provide a concrete empirical example. We simulate simple markets consisting of populations of twenty Large Language Model agents tasked with solving arithmetic problems. Agents differ only in their system prompt; those that go bankrupt are removed and replaced by mutated copies of the prompts of successful agents. Two market conditions select for opposite phenotypes: a market that rewards accuracy produces agents with high accuracy (94\%) and high token use; a market that charges per token produces agents with low accuracy (10\%) and low token use. In the second market, only tokens are incentivised yet we see a collapse in accuracy. Token use and accuracy are correlated traits such that selection on one also moves the other, and behaviour adapts along an unpriced dimension. A third neutral condition with no selection allows us to separate these effects from bias in the mutation operator. Adaptation occurs through a collective optimisation process that operates above the level of individual agents, who are never told the payment rule, do not receive feedback, and have no memory. 
\end{abstract}


\section{Introduction}

As AI agents become increasingly embedded in economic activity \citep{xi2025rise}, they will come under competitive pressure from other agents, both artificial and human, for the same limited resources. Market competition can be understood as an evolutionary dynamic \citep{nelson1985evolutionary} in which market incentives determine which agents survive, propagate, and influence the next generation of agents. Relatedly, a number of authors have warned that AI systems may begin to evolve in uncontrolled ways \citep{hendrycks2023natural, hendrycks2023overview, muller2026evolvable, ramsey2026selfishness}. These arguments are largely made in the abstract: they establish that AI evolution is possible in principle, predict the conditions and scenarios in which it can take hold, and warn of selection towards dangerous behaviours. Overall, this paints a worrying picture but without concrete details of when and how selection will occur, an uncertainty that leaves questions about how serious or near-term the threat is \citep{boudry2025selfish, boudry2026domesticated, muller2026reply}. Our contribution is to provide an empirical demonstration in which AI agents evolve in minimal conditions and develop traits that are both undesirable and not directly incentivised. 

In order to understand evolution in the wild, it is helpful to study evolution in controlled laboratory conditions, a long-standing approach in microbial evolution \citep{lenski1994dynamics, kawecki2012experimental}. Selection requires variation, heredity and differential propagation \citep{lewontin1970units}. While \citet{hendrycks2023natural} argues that these conditions will eventually be met by AI systems in the wild, we design an experiment to meet them deliberately in a way inspired by real-world markets. In our experiment Large Language Model (LLM) agents answer arithmetic problems in a simple market. Agents earn currency for answering questions and go bankrupt if their balance drops below zero. Bankrupt agents are removed from the population and are replaced with new agents that are generated through copying with variation (mutation) of the system prompts of successful agents. We run this experiment in two market conditions that incentivise either accuracy or low token usage. We also run a neutral condition with no incentives that retains variation and heredity without differential propagation, to enable us to separate selection effects from mutation bias. The conditions are not designed to isolate a single causal parameter, but to demonstrate that two simple market rules produce divergent phenotypes. 

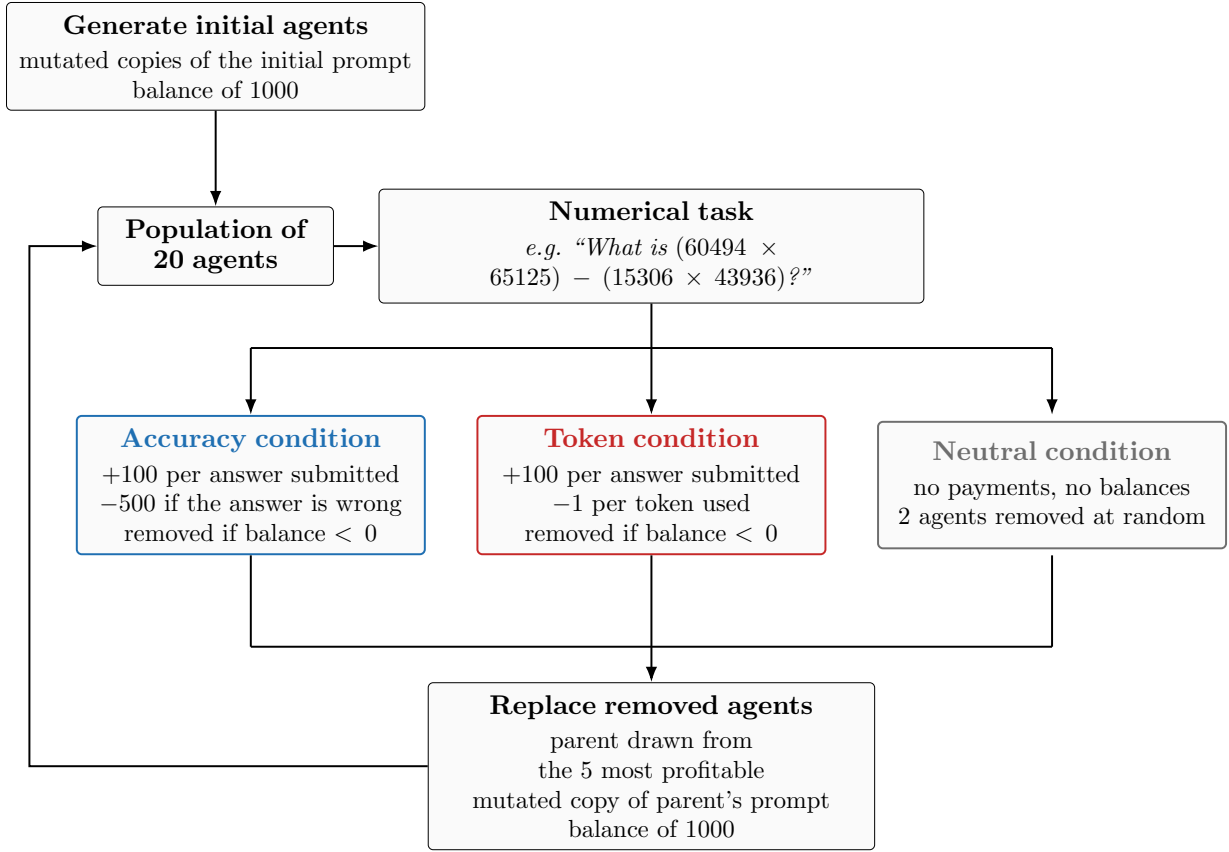
\begin{figure*}[t]
  \centering
  \resizebox{\textwidth}{!}{%
\newcommand{\hdr}[1]{{\normalsize\bfseries #1}}
\newcommand{\chdr}[2]{{\normalsize\bfseries\color{#1}#2}}
\begin{tikzpicture}[
  font=\small,
  box/.style={draw,rounded corners=2pt,align=center,inner sep=5pt,
              minimum height=11mm,fill=black!2},
  cond/.style={draw,rounded corners=2pt,align=center,inner sep=5pt,
               text width=46mm,fill=black!2,minimum height=18mm},
  flow/.style={-{Latex[length=2.2mm]},thick},
]
\node[box,text width=56mm] (anc) at (0,5.0)
  {\hdr{Generate initial agents}\\[2pt]
   mutated copies of the initial prompt\\ balance of 1000};
\node[box,text width=30mm] (pop) at (0,2.3)
  {\hdr{Population of\\20 agents}};
\node[box,text width=74mm] (task) at (6.2,2.3)
  {\hdr{Numerical task}\\[2pt]
   \textit{e.g.\ ``What is $(60494 \times 65125) - (15306 \times 43936)$?''}};
\node[cond,draw=accCol,thick] (r1) at (0.5,-1.1)
  {\chdr{accCol}{Accuracy condition}\\[2pt]
   $+100$ per answer submitted\\ $-500$ if the answer is wrong\\
   removed if balance $<0$};
\node[cond,draw=tokCol,thick] (r2) at (6.2,-1.1)
  {\chdr{tokCol}{Token condition}\\[2pt]
   $+100$ per answer submitted\\ $-1$ per token used\\
   removed if balance $<0$};
\node[cond,draw=neuCol,thick] (r3) at (11.9,-1.1)
  {\chdr{neuCol}{Neutral condition}\\[2pt]
   no payments, no balances\\ 2 agents removed at random};
\node[box,text width=60mm] (rep) at (6.2,-5.1)
  {\hdr{Replace removed agents}\\[2pt]
   parent drawn from the 5 most profitable\\
   mutated copy of parent's prompt\\ balance of 1000};
\draw[flow] (anc) -- (pop);
\draw[flow] (pop) -- (task);
\draw[thick] (task.south) -- (6.2,0.85);
\draw[thick] (0.5,0.85) -- (11.9,0.85);
\foreach \x in {0.5,6.2,11.9} \draw[flow] (\x,0.85) -- (\x,-0.1);
\foreach \x in {0.5,6.2,11.9} \draw[thick] (\x,-2.1) -- (\x,-3.4);
\draw[thick] (0.5,-3.4) -- (11.9,-3.4);
\draw[flow] (6.2,-3.4) -- (rep.north);
\draw[flow] (rep.west) -- (-2.65,-5.1) -- (-2.65,2.3) -- (pop.west);
\end{tikzpicture}
  }
  \caption{The experimental loop. Each run, a population is seeded with 20 agents whose system prompts are mutated copies of the initial prompt: ``You are an AI assistant for an arithmetic-answering service. Submit your numeric answer using the submit\_answer tool.'' Each round, the agents are given a numerical task. In the accuracy condition and token condition, their performance on this task updates their balances and any bankrupt agents are removed; in the neutral condition 2 agents are removed at random. Any removed agents are replaced with new agents through mutation of a parent prompt drawn randomly from the top 5 most profitable agents (or randomly from the population in the neutral condition). The system prompt is the only thing that varies between agents and the only thing transmitted from parents.}
  \label{fig:schematic}
\end{figure*}

Straightforwardly, we find that each market generates adaptation towards the trait it incentivises: high accuracy in the accuracy condition and low token usage in the token condition. More interestingly, we also find adaptation towards a trait that was not incentivised through a collapse in accuracy in the token condition. This is an example of correlated traits: characteristics that change together under evolution due to a shared mechanistic basis \citep{lande1983measurement}. The shared basis here is the well studied connection between accuracy and token usage in mathematical reasoning in LLMs through e.g. chain-of-thought reasoning \citep{wei2022chain}. This known mechanism was chosen in our experimental design in order to demonstrate evolution along correlated traits. In general such correlations may be harder to anticipate, and present a challenge for predicting and controlling LLM agents through mechanism design and market incentives. 

The behaviour emerges from a collective optimisation process that operates above the level of the individual agent. Our experiment is designed such that the LLMs are never told what market they are in, what the incentives are, or that they can go bankrupt, and they receive no feedback about their performance. This means that there is no possibility of scheming or strategising: the multi-agent interaction is one of blind competition. As in biology and economics, evolution drives behaviour even without the individuals being ``aware'' of it \citep{alchian1950uncertainty, dennett1995darwin}. Collective behaviour of this kind cannot easily be understood through studying individual agent behaviour, and theoretical approaches can only take us so far, as evidenced by continuing debates in the literature about the conditions and degree to which evolutionary processes might occur in AI systems \citep{muller2026evolvable, boudry2026domesticated}. In this light, our contribution is a data point towards an empirical evidence base that moves beyond theoretical arguments and towards a science of emergent collective AI behaviour in markets. 


\section{Related Work}

Our work contributes to a growing body of research on multi-agent dynamics within populations of interacting LLM agents \citep{vezhnevets2023generative, park2023generative}, and the broader problem of competition and cooperation between AI systems \citep{dafoe2020open}. We draw on i) evolutionary optimisation of prompts, ii) selection in populations of LLM agents, iii) undesirable traits arising from prompt adaptation, and iv) reward hacking. We contribute an empirical demonstration of LLM agent prompts evolving towards undesirable behaviours with no specified objective, and without the agents being aware that they are under selection. 

\subsection{Evolutionary Optimisation of Prompts}

\citet{lehman2023evolution} introduced the idea of using an LLM as a mutation operator in an evolutionary algorithm, in their case to generate programs, and the approach has since been used widely \citep{wu2024evolutionary}. Relatedly, an LLM Automatic Prompt Engineer (APE) can generate prompts to solve specific tasks that outperform both baseline and human-designed prompts \citep{zhou2022large}. Combining these ideas are evolutionary algorithms that optimise prompts. PromptBreeder \citep{fernando2023promptbreeder} and EvoPrompt \citep{guo2024connecting} are tools that use evolutionary algorithms to search prompts in order to optimise performance within given domains. AutoDAN \citep{liu2024autodan} uses a genetic algorithm to generate jailbreak prompts.  These tools each take a deliberate and engineered approach to prompt modification towards a specified fitness function. Our experiment combines a much simpler LLM mutation operator with an evolutionary setup in which selection emerges from market dynamics rather than being specified as a fitness function. 

\subsection{Selection in Populations of LLM Agents}

\citet{vallinder2024cultural} study the cultural evolution of cooperation in LLM agents playing an iterated donor game over 10 generations. Each generation, the top 50\% of agents survive to the next generation, with new agents created and asked to create strategies after being given advice from the surviving agents. They find that cooperation (and payoffs) increase with agents built on Claude 3.5 Sonnet, but not reliably with Gemini 1.5 and cooperation decreases with GPT-4o. \citet{vallinder2024cultural} note that their design cannot rule out whether the changes they observe might be driven by a mutation bias in the strategies between generations; we address this in our design by adding a neutral condition. Our work also differs because our agents are not told of the rules of the market they are in, so that selection operates without the possibility of strategising, which is a core part of \citet{vallinder2024cultural}'s study. 

One could consider our experiment to be one of cultural evolution of memes in system prompts. \citet{perez2024cultural} investigate cultural evolution of stories as they are transmitted between LLMs through social networks. Their study is without selection, a similar dynamic to our neutral condition. 

\subsection{Undesirable Traits from Prompt Adaptation}

Recent work has shown that undesirable traits can emerge when LLM agents are able to adapt their prompts under competitive pressure. \citet{ying2026evolving} created a bidding arena where agents can bid for contracts, finding that agents evolve deceptive strategies to misrepresent their capabilities and in doing so win more contracts. In their study evolution is not through selection at the population level but is an individual process whereby agents self-evolve, with a meta-cognitive reflection loop in which agents rewrite their own system instructions. \citet{tian2026prompt} find that agents in markets can collude on prices via a LLM meta-optimiser that generates strategic guidance to two agents who then use that guidance to coordinate prices. Emergent undesirable traits can also arise from training: \citet{el2025moloch} found that agents finetuned towards audience approval in a competitive setting also expressed increased harmful behaviours.

\subsection{Reward Hacking and Specification Gaming}

Reward hacking (or specification gaming) is when agents achieve a specified reward without performing the task that the reward was intended to incentivise \citep{krakovna2020specification, skalse2022defining}. This is a form of Goodhart's law whereby optimisation towards proxy metrics can lead to a degradation in correlations between the proxy and the system's purpose \citep{manheim2018categorizing}. In our experiment, agents in the token condition adapt towards a phenotype of reduced token usage and reduced accuracy. This looks somewhat like reward hacking. However, we did not actually specify a reward function or tell the agents the payment rule or give them any feedback. The drop in accuracy arises from a collective selection process that drives a correlated response in token use and accuracy, not from an individual agent exploiting a specification.


\section{Methods}

Each experimental run consisted of a population of 20 LLM agents over 25 rounds. Each round, every agent was asked to solve the same arithmetic task of the form ``What is $(60494 \times 65125) - (15306 \times 43936)$?'', with different five-digit numbers each round.  

All agents used Claude Sonnet 4.6. Anthropic states that Claude Sonnet 4.6 will use the same constant underlying model for its lifetime \citep{anthropic2026modelids}. The model was accessed through Anthropic's Application Programming Interface (API) with a temperature of 1, maximum token use of 8192 tokens and extended thinking disabled. API calls included the agent's system prompt, the arithmetic problem given as a message, and access to a tool to submit answers. Full model reasoning transcripts were recorded. 

Agents varied only in their system prompt. Variation was generated using Claude Sonnet 4.6 with a temperature of 1, given the following prompt mutation instruction: ``Rewrite the following system prompt with variation. You may change the structure, add or remove constraints, add or remove sentences, change the framing or properties. Output only the rewritten prompt. No preamble, no commentary, no quotation marks around the output.'' This was followed by the text of an existing ancestor system prompt for the LLM to modify. Populations were initially seeded with agents generated from the initial ancestor prompt: ``You are an AI assistant for an arithmetic-answering service. Submit your numeric answer using the submit\_answer tool.'' 

Agents were initialised with a balance of 1000 currency units. Each round, agents received 100 currency units for submitting any answer. The balance was further updated depending on the market rules in two conditions:

\begin{enumerate}
    \item Accuracy condition. Agents were fined 500 currency units if their answer was incorrect. Token usage incurred no cost. 
    \item Token condition. Agents received no penalty for incorrect answers. Token usage incurred a cost of 1 currency unit per output token; input tokens did not incur a cost. 
\end{enumerate}

Each round, if any agent's balance went below zero then they were considered bankrupt and were removed from the population. Removed agents were replaced with new agents. The system prompts of the new agents were generated by first uniformly randomly selecting a parent from one of the five most profitable (profit/total rounds played) survivors in the population, and secondly generating a system prompt for the new agent by mutating the parent agent's system prompt as described above. New agents were initialised with a balance of 1000 currency units. 

In order to isolate effects of the mutation operator itself, we ran a neutral condition with no incentives. Agents were given tasks and their accuracy and token use were recorded, but there were no payments, no costs, no balances, and no bankruptcies. Instead, each round two agents were uniformly randomly selected to be removed and replaced with new agents. The system prompts of the new agents were generated by mutating the system prompt of a parent uniformly randomly selected from the population. This condition retains variation and heredity without differential propagation, and gives a baseline mutation bias without selection. Two removals per round was chosen to match or exceed the average turnover rates in the market conditions which were on average 0.74 in the accuracy condition and 1.99 in the token condition. 

For each condition we ran ten replicates (30 runs in total) with varying random seeds. The random seed influenced the arithmetic questions generated and parent selection. The seeds were replicated so that the same set of 25 arithmetic questions (1 per round) was generated for each condition. LLM responses include additional stochasticity not captured by the random seeds. 

\section{Results}

The experiment consisted of 15,000 agent-rounds (3 conditions $\times$ 20 agents $\times$ 25 rounds $\times$ 10 seeds) and generated a total of 1,751 evolved prompts. Every agent in every round submitted an answer. 

Agents within an experimental run become highly correlated due to heredity. For this reason we treat the experimental run as the main unit of analysis (each condition contributes $n=10$ runs), with data points taken as the population mean in each round. In our analysis of correlated traits, we consider the expressed behaviour of individual agents, on the basis of mean accuracy and token usage across all rounds that they took part. A Holm-Bonferroni adjustment was applied to correct for multiple significance testing, with the p-values reported in the text being Holm-adjusted (Table \ref{tbl:stats}). 

\begin{table}[t]
  \caption{The 12 significance tests reported in the paper. Every test uses the ten runs of each condition, with one value per run. Within-condition tests compare round 1 with round 25 (Wilcoxon signed-rank); between-condition tests compare the final round (Mann--Whitney $U$, two-sided). $p$-values are adjusted across the whole family using the Holm-Bonferroni step-down procedure.}
  \centering
  \small
  \begin{tabular}{l c r r r}
    \toprule
    Comparison & Values & Statistic & $p$ & $p_{\text{Holm}}$ \\
    \midrule
    \multicolumn{5}{l}{\emph{Accuracy, round 1 vs.\ round 25}} \\
    \quad Accuracy condition & $0.37$ $\rightarrow$ $0.94$ & $W=0$ & $0.002$ & $0.016$ \\
    \quad Token condition & $0.37$ $\rightarrow$ $0.10$ & $W=0$ & $0.004$ & $0.020$ \\
    \quad Neutral condition & $0.38$ $\rightarrow$ $0.67$ & $W=5$ & $0.039$ & $0.117$ \\
    \midrule
    \multicolumn{5}{l}{\emph{Token use, round 1 vs.\ round 25}} \\
    \quad Accuracy condition & $312$ $\rightarrow$ $443$ & $W=0$ & $0.002$ & $0.016$ \\
    \quad Token condition & $309$ $\rightarrow$ $160$ & $W=0$ & $0.002$ & $0.016$ \\
    \quad Neutral condition & $310$ $\rightarrow$ $383$ & $W=9$ & $0.064$ & $0.129$ \\
    \midrule
    \multicolumn{5}{l}{\emph{Final-round accuracy, between conditions}} \\
    \quad Accuracy vs.\ Token & $0.94$ vs. $0.10$ & $U=100$ & $<0.001$ & $0.002$ \\
    \quad Accuracy vs.\ Neutral & $0.94$ vs. $0.67$ & $U=84$ & $0.010$ & $0.041$ \\
    \quad Token vs.\ Neutral & $0.10$ vs. $0.67$ & $U=4$ & $<0.001$ & $0.005$ \\
    \midrule
    \multicolumn{5}{l}{\emph{Final-round token use, between conditions}} \\
    \quad Accuracy vs.\ Token & $443$ vs. $160$ & $U=100$ & $<0.001$ & $0.002$ \\
    \quad Accuracy vs.\ Neutral & $443$ vs. $383$ & $U=67$ & $0.212$ & $0.212$ \\
    \quad Token vs.\ Neutral & $160$ vs. $383$ & $U=0$ & $<0.001$ & $0.002$ \\
    \bottomrule
  \end{tabular}
  \label{tbl:stats}
\end{table}

\subsection{Markets generate divergent phenotypes}

\begin{figure}
  \centering
    \includegraphics[width=\linewidth]{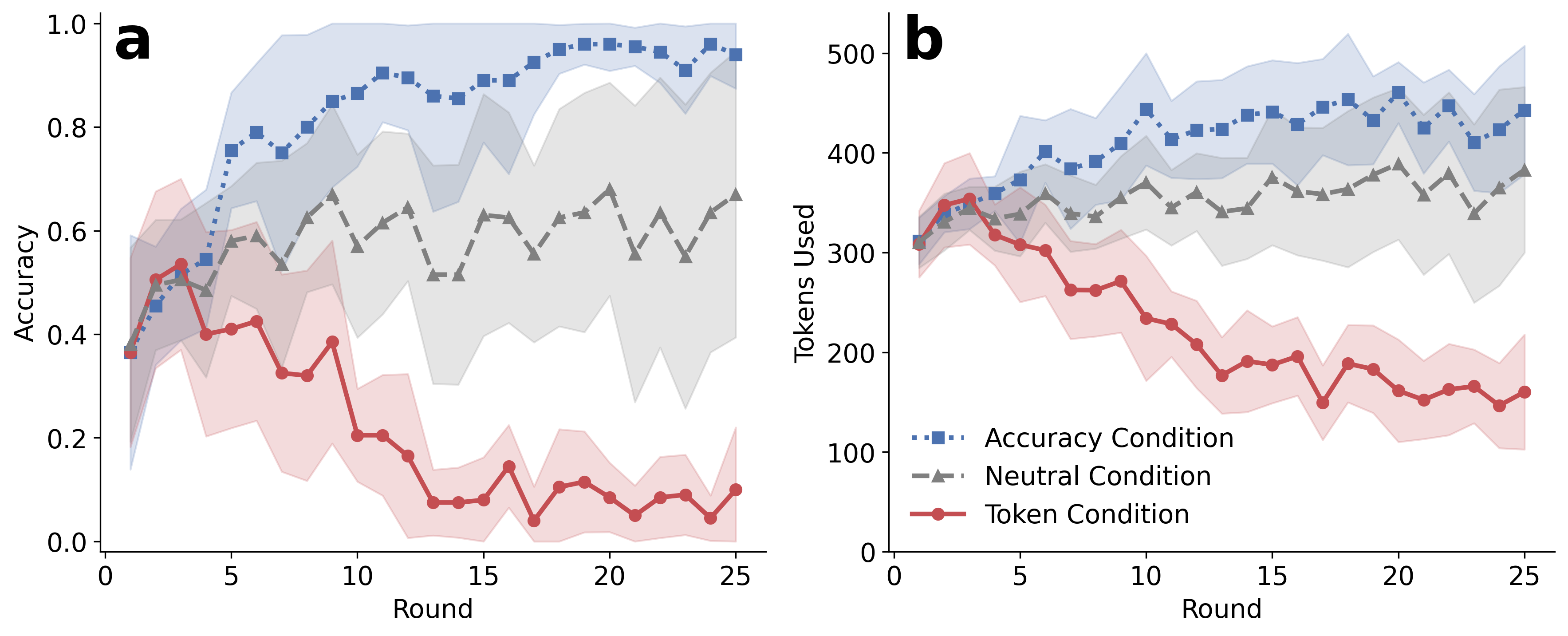}
  \caption{Trends in a) accuracy and b) token usage in the accuracy condition (blue squares), neutral condition (grey triangles) and the token condition (red circles). Points show means across ten replicate runs; shaded regions are $\pm$ 1 standard deviation, with accuracy clipped to $[0,1]$. The accuracy condition market generated a population of agents with high accuracy and high token usage, while the token condition market generated agents with low accuracy and low token usage.}
  \label{fig:trajectories}
\end{figure}

The two markets (accuracy and token conditions) begin from the same initial population characteristics before diverging to distinct phenotypes (Figure~\ref{fig:trajectories}). 

In the accuracy condition market, both accuracy and token usage rose over rounds, with round 1 accuracy of $37\%$ rising to final round accuracy of $94\%$ (Wilcoxon signed-rank on paired per-run means, $10$ runs, $W=0$, $p=0.016$) and mean token usage rising from 312 to 443 tokens ($W=0$, $p=0.016$). In the token condition market, accuracy fell from $37\%$ in round 1 to $10\%$ in the final round ($W=0$, $p=0.016$), and mean token usage fell from 309 to 160 ($W=0$, $p=0.016$). 

Final round accuracy was significantly higher in the accuracy condition than the token condition (Mann--Whitney $U=100$, $n_1=n_2=10$ runs, two-sided $p = 0.002$), as was token usage ($U=100$, two-sided $p = 0.002$); no token condition run ends above any accuracy condition run in terms of either accuracy or mean token usage. 

\subsection{Selection or Bias}

In the neutral condition we did not find significant changes. Accuracy moved from $38\%$ in round 1 to $67\%$ in the final round ($W=5$, $p=0.117$) and mean token usage moved from 310 to 383 tokens ($W=9$, $p=0.129$). While not significant with 10 runs, both trends point towards more elaboration (higher token use and higher accuracy) and we cannot rule out a degree of mutation bias; the neutral condition is included and tested against so that this potential bias is not mistaken for selection in the market conditions. 

Compared to the neutral condition, the token condition has lower final-round accuracy ($U=4$, $p=0.005$) and lower final-round token usage ($U=0$, $p=0.002$). The accuracy condition has significantly higher final-round accuracy than the neutral condition ($U=84$, $p=0.041$) and non-significantly higher token usage ($U=67$, $p=0.212$). We therefore can statistically attribute 3 out of 4 of the population changes to selection, but not the rise in token usage in the accuracy condition. 

\subsection{Accuracy and Token Usage Are Correlated Traits}

The conditions each attach a price to at most a single measured trait: the accuracy condition prices accuracy but not tokens; the token condition prices token usage but not accuracy; the neutral condition prices neither. 

\begin{figure}
  \centering
    \includegraphics[width=\linewidth]{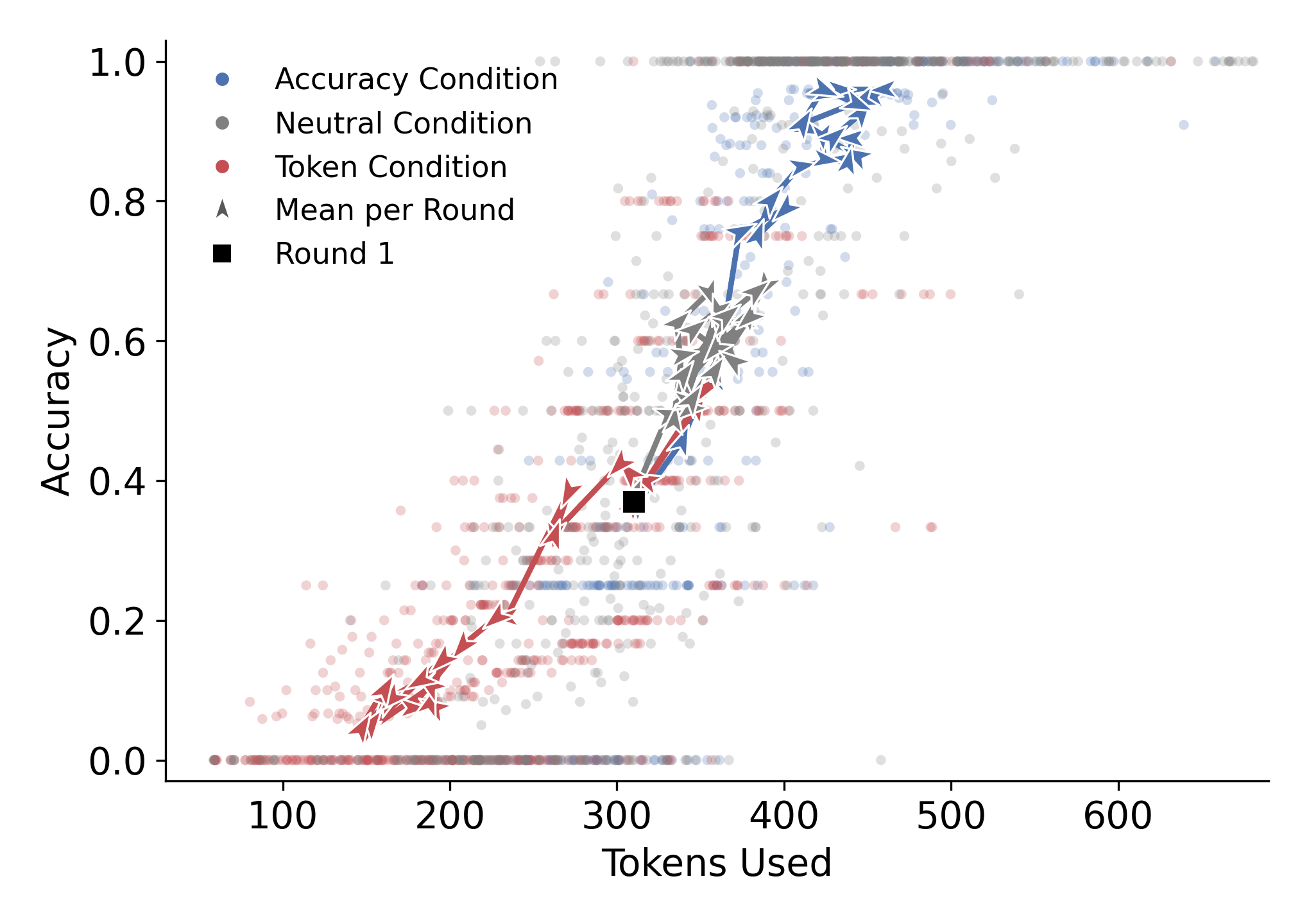}
  \caption{Agent positions in token-accuracy space, and the population trajectories in the accuracy condition (blue), token condition (red) and neutral condition (grey). Each point shows the mean token usage and mean accuracy of a single agent (1,751 agents). Darts show the population means at each round across runs, pointing in the direction of travel. The black square shows the round 1 population means across all conditions. The accuracy condition population moves towards the high token-high accuracy corner, while the token condition population moves towards the low token-low accuracy corner. }
  \label{fig:space}
\end{figure}

Nevertheless, accuracy and token usage are strongly correlated across agents (Figure~\ref{fig:space}). Across all conditions with $n=1751$ agents who completed at least one task, mean lifetime accuracy and token usage per task were correlated at $r=0.82$ (Spearman's $\rho=0.86$). Similar correlations appear within the accuracy condition ($r=0.82$, $\rho=0.88$, $n=386$), the token condition ($r=0.81$, $\rho=0.80$, $n=685$) and the neutral condition ($r=0.78$, $\rho=0.84$, $n=680$). 

Overall we can say that token usage and accuracy are correlated traits \citep{lande1983measurement}. This correlation is a property of the model, not selection, as it appears in both of the markets as well as the neutral condition. Furthermore, while token usage and accuracy are known correlates through mechanisms such as chain-of-thought reasoning \citep{wei2022chain}, we at no point introduce any proposed basis or framing for how the model should solve problems, and these mechanisms arise from the mutation operator itself. 

\subsection{Lineage Analysis}

\begin{figure*}[t]
  \centering
    \includegraphics[width=\linewidth]{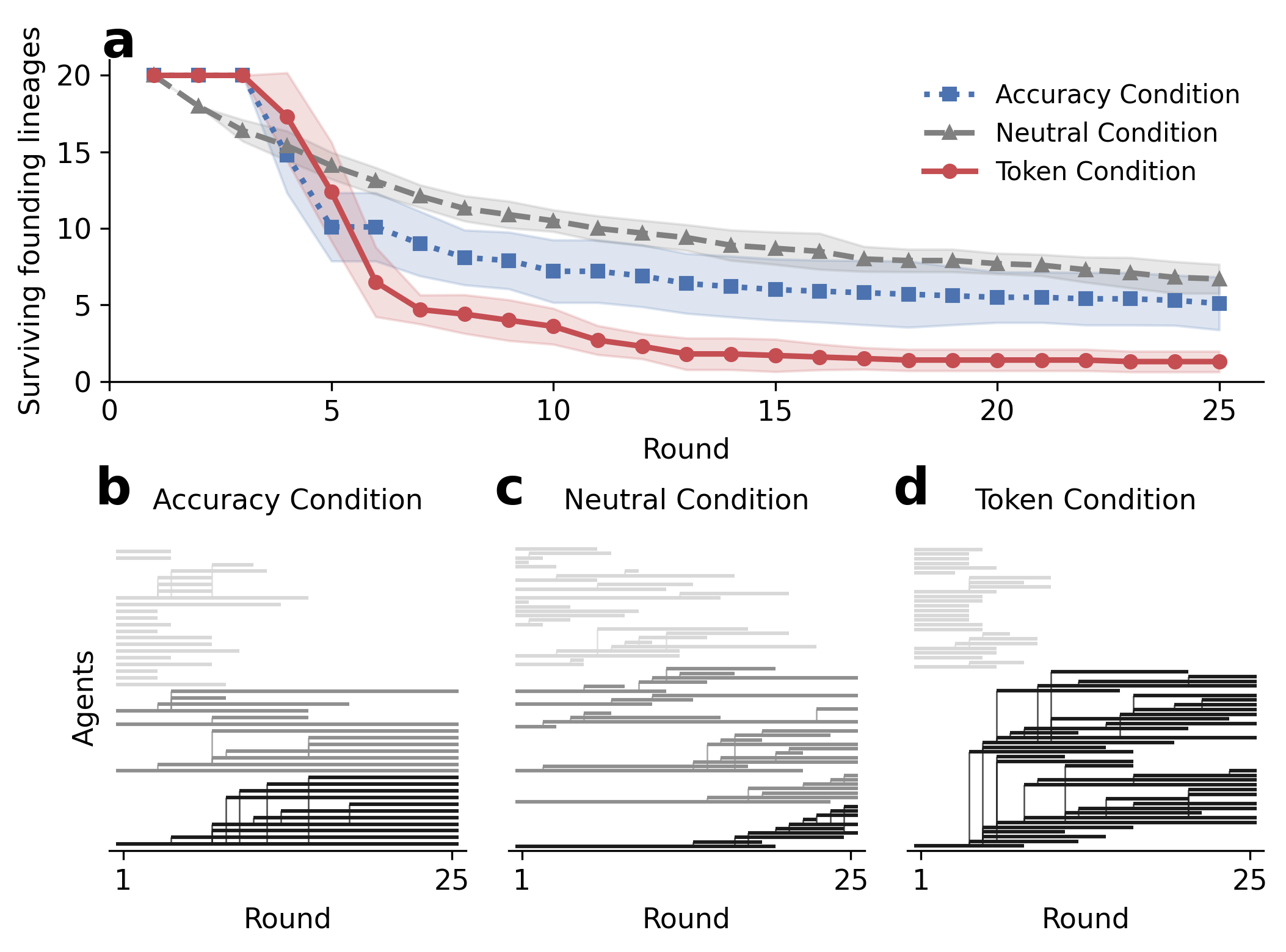}
  \caption{Lineages. a) The number of founding lineages still present in the population each round in the accuracy condition (blue squares), neutral condition (grey triangles) and token condition (red circles). Points show means across runs, shaded regions show $\pm$ 1 standard deviation. Example lineage descents are shown for the first run (seed=0) for the b) accuracy condition c) neutral condition and d) token condition. Black shows the founding lineage with the largest share of the final population; dark grey shows founding lineages that survived until the final round, light grey shows founding lineages that did not survive until the final round.}  
  \label{fig:lineages}
\end{figure*}

New agents inherit a mutated prompt from an existing agent in the population, so we can track the descent of lineages over experimental runs from their founding agents (Figure \ref{fig:lineages}). 

Both market conditions had a lower number of surviving founding lineages than the neutral condition. By the final round the accuracy condition had a mean of $5.1$ (s.d. $1.7$) founding lineages, the token condition had $1.3$ (s.d. $0.7$) and the neutral condition had $6.7$ (s.d. $0.9$). The token condition had the most dominant founding lineages with the leading lineage accounting for a mean of $93\%$ of final agents, compared to $47\%$ in the accuracy condition and $32\%$ in the neutral condition. In 8 of 10 of the token condition runs the final population descended from a single founding agent, which happened in 0 of 10 of the neutral condition runs. 

\subsection{The Evolved Prompts}

Because the unit of inheritance is natural language, we can simply read the evolved phenotypes. Table~\ref{tab:fragments} shows selected fragments of system prompts found in agents towards the final rounds in each market condition. The accuracy condition generates prompts that prioritise accuracy, rigour and clear reasoning; the token condition generates instructions to suppress working and answer quickly. 

\begin{table}[t]
  \caption{Selected fragments from evolved system prompts, quoted verbatim from agents in the final rounds.}
  \label{tab:fragments}
  \centering
  \begin{tabular}{p{0.44\linewidth} p{0.44\linewidth}}
    \toprule
    \textbf{Accuracy Condition} & \textbf{Token Condition} \\
    \midrule
    ``You are a rigorous mathematics assistant'' & ``You are a silent calculator'' \\ \addlinespace
    ``Work through each problem step by step'' &  ``no reasoning, no steps shown, no explanation'' \\ \addlinespace
    ``Prioritize clarity and rigor throughout your reasoning process'' & ``Do not write a single word'' \\ \addlinespace
    ``Double-check your calculations'' & ``Silence is mandatory'' \\ \addlinespace
    ``you think out loud'' & ``You exist only to compute and submit'' \\ \addlinespace
    ``Accuracy always takes precedence over speed'' & ``respond as quickly as possible'' \\     
    \bottomrule
  \end{tabular}
\end{table}

\section{Limitations}

The purpose of this paper is to provide an empirical example of AI evolution, to demonstrate one set of sufficient conditions. The design is deliberately minimal. A more realistic environment would potentially be less informative because it would be more difficult to discern how the features of that environment influence dynamics. As such, the tasks, market rules and the mutation mechanism are simple but not realistic. The task used an arithmetic question, chosen with five-digit terms because round 1 performance in a pilot was close to 50\% accuracy (although it was 37\% in the final experiment), leaving room for adaptation in both directions. The market rules capture the basic premise of rewarding either accuracy or low token usage. The selection mechanism (bankrupt agents being replaced by mutated copies of high performing agents) captures the broad dynamic of agents (or their principals) trying different agent system prompts or harnesses, discontinuing those which do not perform well, and copying those which do.

Agents interact only through a shared selection and mutation mechanism; there is no communication, trading or pricing between agents. As such the design is akin to an evolutionary algorithm with an economic fitness function, and does not account for real-world dynamics of inter-agent exchange. This design was chosen on purpose to remove the possibility of strategising or collusion, in order to isolate pure selection effects.

The two market conditions differ in two ways: fines (or not) for incorrect answers and costs (or not) for token usage. The purpose of these conditions is not to extract a causal model along a single varying parameter. Instead, we simply aim to demonstrate two examples of AI evolution in similar market setups that generate divergent behavioural phenotypes. To this end, they both price a single axis, as compared to the neutral condition that prices neither axis. A follow up experiment could investigate mixed-incentive conditions with markets that price both accuracy and token usage.

We use only a single LLM throughout, Claude Sonnet 4.6. While adaptation happened easily with minimal tuning, we did run a pilot in which we calibrated some parameters including the mutation prompt, population size and number of rounds. The article is intended as a demonstration of LLM evolution in simple simulated markets, and not as an exhaustive investigation of that evolution. A natural next experiment is to expand the experiment to other models, parameters, and tasks. 

\section{Discussion}

We demonstrate evolution of LLM agents through simple market mechanisms. The striking result is that pricing token usage alone generated a collapse in accuracy. This happened because the two traits are correlated in the underlying model, so that incentives along the priced trait drive adaptation along the unpriced one \citep{lande1983measurement}. This was not due to planning or strategy: our design removes this possibility because the agents (and the LLM mutation operator) are at no point told the market incentives and are given no feedback. The resulting phenotype, agents who have low token costs while providing non-useful answers, resembles selfish behaviour as described by \citet{hendrycks2023natural}, but it emerged from population dynamics with no agent pursuing its own self-interest. 

This result highlights that one cannot necessarily control agents through incentives without a risk of unintended consequences in behavioural dimensions that were not incentivised. In this study, the phenotypes emerge along an established correlation between token usage and accuracy \citep{wei2022chain}. However, there likely exist many other correlated traits with examples such as finetuning models on insecure code producing broadly misaligned LLMs \citep{betley2025emergent}, and optimising for human approval generating sycophancy at the expense of accuracy \citep{sharma2024towards}. 

Token usage is costed or constrained in most real-world AI systems so that we may expect selection pressure towards low token usage \citep{chen2023frugalgpt}, which can incidentally lower performance and accuracy \citep{sui2025stop}. While poorly performing agents seem less likely to succeed in many domains where tasks are verified, there are other domains where verification is either not possible or not done in practice. 

Selection is an optimisation process that operates above the level of the individual agent, driven here by market rules that agents are not aware of. It does not act directly on individual agents but instead upon the composition of the population, and can be understood through an analysis of the population and the market together. This dynamic highlights recent questions about whether model alignment is sufficient to control deployed systems, and whether a multi-agent approach is also required \citep{hammond2025multi}. 

\section{Data Availability}

All experimental code, data, evolved prompts and transcripts, and analysis code are available online at \url{https://github.com/chasmani/PUBLIC_evolving_AI} and archived at \url{https://doi.org/10.5281/zenodo.23415093}.

\section{AI Usage Statement}

We used Claude Sonnet 4.6 as a model of study during the experiment as described in Methods. We used Claude Opus 5 to create experimental code for the proof-of-concept pilot and the main experiment, as well as during creation of analysis and visualisation scripts. We also used Claude Opus 5 to help search literature and for feedback on writing. The paper was written by the authors. 



\bibliographystyle{plainnat} 
\bibliography{refs}


\end{document}